%% file: main_edit1_clean.tex
\documentclass[twocolumn]{aastex631}

\newcommand\rp{\emph{r}-process}

\usepackage{enumerate}
\usepackage{graphicx}
\usepackage{amsmath}
\usepackage{float}
\usepackage{xcolor}
\usepackage{hyperref}
\hypersetup{
    bookmarks=true,                 
    unicode=false,                  
    pdftoolbar=true,                
    pdfmenubar=true,                
    pdffitwindow=true,              
    pdfstartview={FitH},            
    pdftitle={Superheavy Elements in Kilonovae}, 
    pdfauthor={eholmbeck},          
    pdfsubject={Astrophysics},    
    pdfcreator={dvipdf},            
    pdfproducer={dvipdf},           
    pdfkeywords={r-process, stars}, 
    pdfnewwindow=true,              
    colorlinks=true,                
    linkcolor=magenta,              
    citecolor=teal,                 
    filecolor=magenta,              
    urlcolor=cyan,                  
    breaklinks=true,
    linktocpage
}

\graphicspath{{./}{Figures}}
\newcommand\Msun{M$_\sun$}

\received{April 6, 2023}
\revised{May 18, 2023}
\accepted{May 29, 2023}
\submitjournal{\apjl}

\shorttitle{Superheavy Elements in Kilonovae}
\shortauthors{Holmbeck et al.}

\begin{document}

\title{Superheavy Elements in Kilonovae}

\correspondingauthor{Erika M.\ Holmbeck}
\email{eholmbeck@carnegiescience.edu}

\author[0000-0002-5463-6800]{Erika M.\ Holmbeck}
\altaffiliation{NHFP Hubble Fellow}
\affiliation{Observatories of the Carnegie Institution for Science, 813 Santa Barbara St., Pasadena, CA 91101, USA}

\author[0000-0003-3340-4784]{Jennifer Barnes}
\affiliation{Kavli Institute for Theoretical Physics, Kohn Hall, University of California, Santa Barbara, CA 93106, USA}

\author[0000-0003-0031-1397]{Kelsey A.\ Lund}
\affiliation{Department of Physics, North Carolina State University, Raleigh, NC 27695 USA}
\affiliation{Theoretical Division, Los Alamos National Laboratory, Los Alamos, NM 87545, USA}
\affiliation{Center for Nonlinear Studies, Los Alamos National Laboratory, Los Alamos, NM 87545, USA}

\author[0000-0002-4375-4369]{Trevor M.\ Sprouse}
\affiliation{Theoretical Division, Los Alamos National Laboratory, Los Alamos, NM 87545, USA}
\affiliation{Center for Theoretical Astrophysics, Los Alamos National Laboratory, Los Alamos, NM 87545, USA}

\author[0000-0001-6811-6657]{G.\ C.\ McLaughlin}
\affiliation{Department of Physics, North Carolina State University, Raleigh, NC 27695 USA}

\author[0000-0002-9950-9688]{Matthew R.\ Mumpower}
\affiliation{Theoretical Division, Los Alamos National Laboratory, Los Alamos, NM 87545, USA}
\affiliation{Center for Theoretical Astrophysics, Los Alamos National Laboratory, Los Alamos, NM 87545, USA}

\begin{abstract}
As LIGO-Virgo-KAGRA enters its fourth observing run, a new opportunity to search for electromagnetic counterparts of compact object mergers will also begin.
The light curves and spectra from the first ``kilonova" associated with a binary neutron star binary (NSM) suggests that these sites are hosts of the rapid neutron capture (``\emph{r}") process.
However, it is unknown just how robust elemental production can be in mergers.
Identifying signposts of the production of particular nuclei is critical for fully understanding merger-driven heavy-element synthesis.
In this study, we investigate the properties of very neutron rich nuclei
for which superheavy elements ($Z\geq 104$) can be produced in NSMs and whether they can similarly imprint a unique signature on kilonova light-curve evolution.
A superheavy-element signature in kilonovae represents a route to establishing a lower limit on  heavy-element production in NSMs as well as possibly being the first evidence of superheavy element synthesis in nature.
Favorable NSMs conditions yield a mass fraction of superheavy-elements is $X_{Z\geq 104}\approx 3\times 10^{-2}$ at 7.5 hours post-merger.
With this mass fraction of superheavy elements, we find that the component of kilonova light curves possibly containing superheavy elements may appear similar to those arising from lanthanide-poor ejecta.
Therefore, photometric characterizations of superheavy-element rich kilonova may possibly misidentify them as lanthanide-poor events.
\end{abstract}



\section{Introduction}
\label{sec:intro}

The joint multi-messenger detection of gravitational waves and electromagnetic emission from the merger of two neutron stars in 2017 \citep{abbott2017b} represented a new opportunity for observational studies of heavy-element production.
Sixty years after the \rp\ was first theorized \citep{Burbidge1957,Cameron1957}, the NSM event GW170817 \citep{abbott2017} and its corresponding afterglow AT2017gfo \citep{coulter2017,cowperthwaite2017,drout2017,kilpatrick2017,shappee2017} provided the first direct evidence that binary neutron star mergers can host the \rp\ \citep{kasen2017,Smartt2017_gw170817,tanaka2017_gw170817,tanvir2017_gw170817,Villar2017,watson.ea_knSr}, supporting a theory proposed decades earlier \citep{Lattimer1974}.
Consequently, recent studies of compact object mergers and \rp\ nucleosynthesis have focused on macrophysical and microphysical properties that can be deduced from compact object merger light curves, e.g., broad parameter estimation of ejecta masses and velocities \citep{Coughlin2019,Radice2019,Breschi2021,Heinzel2021}. In addition, fission, which was known to be a critical factor in determining the abundance pattern, e.g. \citet{Beun:2007wf,Vassh:2019cey}, was now realized to be an important predictor of the light curve.
Recent studies have investigated the detailed elemental compositions of the ejecta and fission properties of heavy nuclei \citep{Wu2019,Vieira2023}.
In general, composition inference and the claim that \rp\ elements were synthesized in the GW170817 event are based on the effect of high-opacity lanthanides ($57\leq Z\leq 71$) on the light-curve evolution, and it cannot currently be definitively claimed that anything beyond the lanthanides were created in that event, see \citet{Zhu2018}.

Recent works have detailed the effects of ejecta mass and composition on kilonova light curves \citep{barnes2021,zhu2021,lund2023} and have even identified an individual nucleus that can power the light curve at late times: $^{254}$Cf \citep{Zhu2018}.
With a half life of about 60 days, the energy released by the spontaneous fission of $^{254}$Cf, can prolong the evolution of light in the \emph{JHK} bands, leaving a measurable excess of light beginning as early as 25 days post-merger.
Such a late-time detection would be observational proof that the actinides ($89\leq Z\leq 103$) are synthesized in mergers---not just the lanthanides.
Other works have also discussed individual nuclei that can power the late-time light curves \citep[e.g.,][]{Wu2019}.

An earlier signal could possibly arise from the energetic decay of even heavier nuclei: the ``superheavy" elements with $Z\geq 104$.
Whether these elements are even produced by nature is a topic of debate and depends sensitively on the nuclear physics at high proton and neutron numbers \citep{Holmbeck2023}. 
In this work, we explore a subset of nuclear models that can produce superheavy elements in an NSM environment and study the effect that superheavy elements have on kilonova light curves.

\section{Nucleosynthesis of superheavy elements}

To investigate the effect of superheavy elements on kilonova light curves, we first choose conditions that favorably produce these elements.
Across \rp\ nucleosynthesis studies, see \citep{Arcones:2022jer} and references therein, it is found that conditions that generally produce high actinide abundances are the low-entropy and high neutron-richness of dynamical ejecta from NSMs.
Therefore, we choose a single dynamical ejecta trajectory from the 1.4--1.4\,\Msun\ NSM simulations of S.\ Rosswog \citep{piran2013,rosswog2013} as in \citet{korobkin2012}.
We run nucleosynthesis simulations with the nuclear reaction network code Portable Routines for Integrated nucleoSynthesis Modeling \citep[PRISM;][]{Sprouse2020} and begin the network in nuclear statistical equilibrium at 10 GK.

Even under the most favorable environmental conditions (low-entropy and high neutron-richness), superheavy-element production is not guaranteed.
Specifically, low fission barriers can cause heavy nuclei to fission before superheavy elements can be produced \citep[see, e.g.,][]{Mumpower2018, Vassh2019}.
To investigate the range of superheavy-element production that can be achieved by different nuclear models, we test three nuclear model and fission barrier combinations: the Finite Range Liquid Droplet Model \citep[FRLDM;][]{Moller2015} for the 2012 version of the Finite Range Droplet Model \citep[FRDM;][]{moller2012,Moller2016}, fission barriers based on the Koura-Tachibana-Uno-Yamada model \citep[KTUY][]{Koura2005} for Duflo-Zuker \citep[DZ;][]{Duflo1995}, and Hartree-Fock-Bogoliubov \citep[HFB;][]{Goriely2007} for HFB \citep[version 27;][]{Goriely2013}.
Experimentally measured data are used wherever possible in all nucleosynthesis calculations, regardless of the other choice of theoretical model.
Otherwise, reaction and decay rates are calculated as self-consistently as possible with the given nuclear model (or experimentally measured mass, if available, as in \citealt{Mumpower2015}) and fission barrier heights.
Importantly, all data regarding the superheavy elements involved in \rp\ calculations are entirely calculated from theory, as known superheavy-element data is restricted to less neutron-rich isotopes.

	\begin{figure}[t]
	\centering
	\includegraphics[width=\columnwidth]{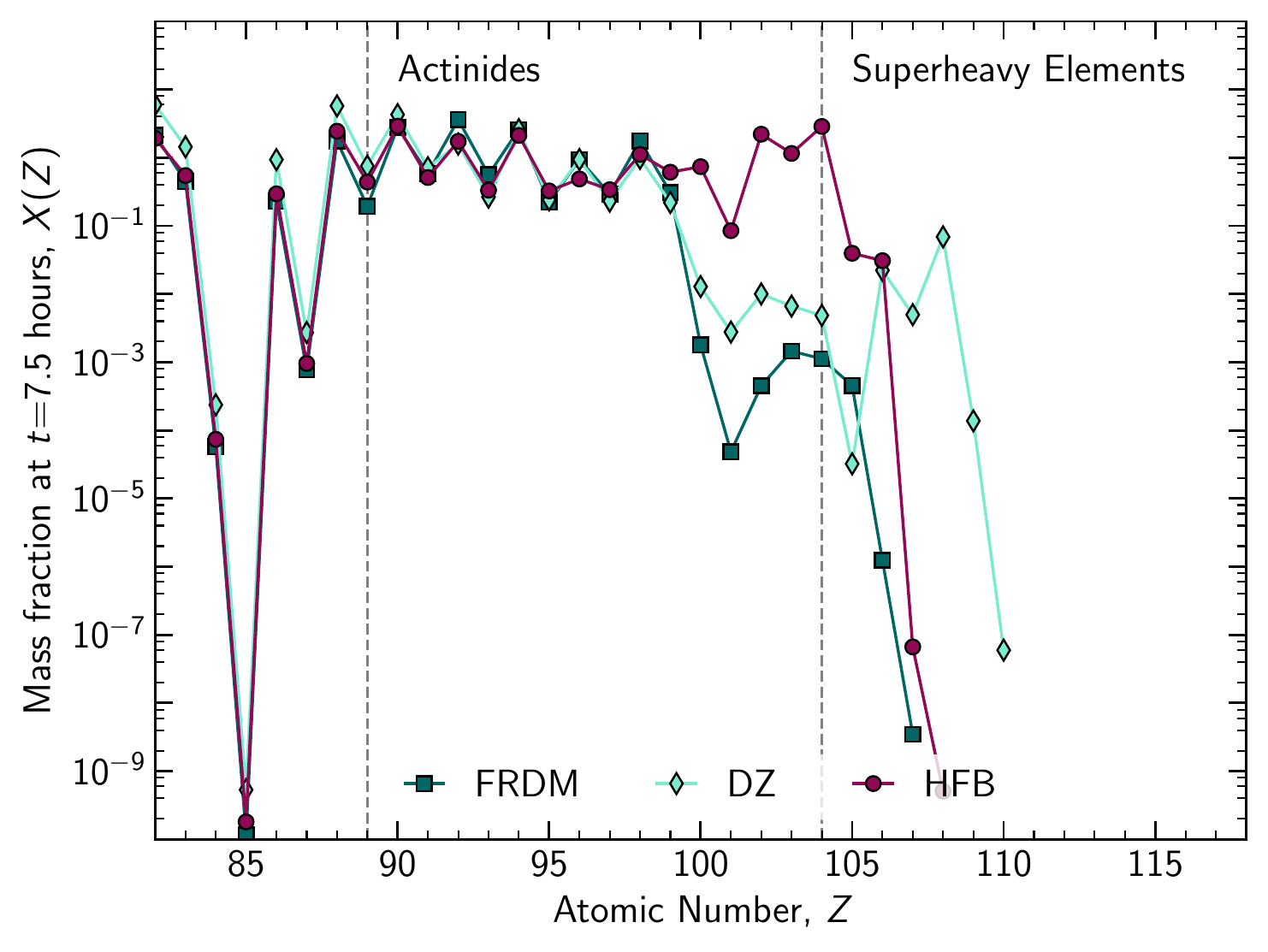}
	\caption{Nucleosynthesis network mass-fraction yields of the heaviest elements at 7.5 hours post-merger for three nuclear mass model and fission barrier combinations. Under the same astrophysical conditions, there can be over three orders of magnitude difference in the mass fractions of superheavy elements that are synthesized, arising solely from what nuclear models are adopted.
	}
	\label{fig:massfracs}
	\end{figure}

We use the same astrophysical conditions for all three cases. 
As the composition evolves over time, nuclear reheating from the decay of radioactive species affects the temperature evolution of the trajectory.
The reaction rates are in turn affected by the temperature change relative to the original trajectory. 
Since the energy released by radioactive nuclei depends on nuclear mass differences, the reheating calculation is performed to be consistent with the specific theoretical (and experiment) nuclear masses implemented.
We adjust the temperature of the trajectory from nuclear reheating self-consistently with the nuclear physics input to determine the extent to which the temperature evolution needs modification and recalculate reaction rates accordingly.

Figure~\ref{fig:massfracs} shows the heavy-element mass fractions ($X(Z)=\sum_i A_iY_i$, where $i$ is every isotope of element $Z$, and $Y$ is the number abundance) at 7.5 hours post-merger for three nuclear variations.
The HFB model has the highest mass fraction of superheavy elements at this time, concentrated primarily at $Z=104$ (Rf).
FRDM is among the most pessimistic, producing superheavy elements of about four orders of magnitude less than the actinides, while DZ achieves significant mass fractions as heavy as $Z=108$ (Hs), though in overall less amounts than HFB.
For this study, we will choose the two extremes, FRDM and HFB, to examine the effect that superheavy elements have on NSM light curves.
At 7.5 hours, the mass fractions of lanthanides, actinides, and superheavy elements for FRDM (HFB) are $X_{\rm lan}=0.094$ (0.10), $X_{\rm act}=0.14$ (0.18), and $X_{Z\geq 104}=1.6\times 10^{-5}$ ($3.0\times 10^{-2}$), respectively.
The mass fractions of the superheavy nuclei at 7.5 hours, 1 day, 1 week, and 1 month are given in the Appendix.

\section{Light curve calculation}

The two most basic ingredients to translate abundance evolution into light curves are heating (including thermalization effects) and ejecta opacities.
First, as radioactive nuclei made in the \emph{r}-process decay, they release energy, a fraction of which is converted to thermal photons (``thermalized"), which ultimately power the kilonova's electromagnetic emission.

\subsection{Heating rates}

The output of PRISM, along with the underlying nuclear data---both experimental and calculated---are used to determine the rate at which heat is generated by the ejecta as well as its composition over time, as in \citet{zhu2021}.
We calculate light curves representing three different levels of contribution from heavy-element decay.
First we calculate heating rates and light curves for all nuclei from the PRISM output (``$Z$ = all").
Secondly, as a simple test, we remove the effects $Z\geq 104$ nuclei (the superheavy elements) from the heating-rate calculation (``$Z<104$").
In other words, any reaction or decay involving $Z\geq 104$ elements is assumed to contribute zero energy, but we do not otherwise remove the nuclei from the network calculation.
Lastly, we mute the effects of all $Z\geq 90$ nuclei on the heating rate calculation, such that only $Z\leq 89$ elements impact the heating of the system (``$Z<90$").

For the sake of studying the direct effect that superheavy elements have on kilonova light curves, we use an identical astrophysical environment (which is subject to nuclear-model-dependent reheating from actinides and superheavy elements) and alter the heating rate post-facto, as described above, though we note that a more complex approach is necessary to account for the indirect effects that superheavy elements would have on the light curve, e.g., through the decay of their fission products at lower atomic masses.
We expect that removing both the direct and indirect products of superheavy elements would only further differentiate the heating rate, and thus our approach provides a lower bound.

	\begin{figure}[t]
	\centering
	\includegraphics[width=\columnwidth]{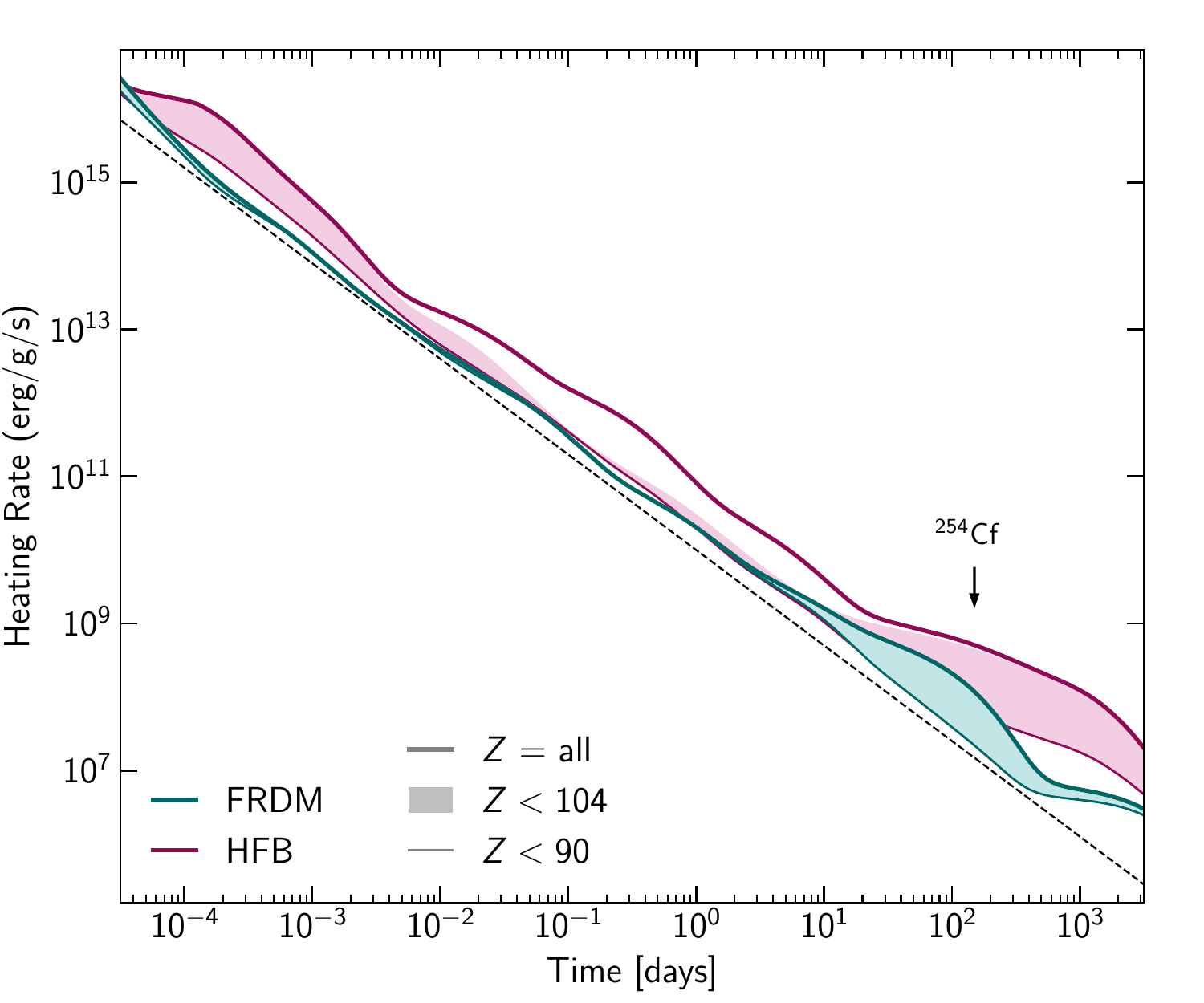}
	\caption{Heating rates over time for two extremes of nuclear variations showing the contribution by all elements (thick, solid lines), only up to the actinides (shaded), and excluding the actinides and superheavy elements (thin, solid lines). The dashed line is a simple model in which the heating rate evolves scales with time by $10^{-1.3}$ \citep[see, e.g.,][]{Metzger2019}. The divergence between the shaded region and the thick lines (e.g., at about $10^{-2}$--$10^{1}$ days for HFB) indicates heating uniquely from the superheavy elements.
	}
	\label{fig:heating}
	\end{figure}

Figure \ref{fig:heating} shows the total heating rates for the three nuclear models for three different cases: with only $Z<90$ nuclei (thin lines), only $Z<104$ nuclei (filled), and all nuclei (solid lines).
Going from the thin lines to the top of the filled regions shows the direct effect that actinides have on the heating rates.
Notably, both models show a dramatic increase at about 100 days due to the spontaneous fission of $^{254}$Cf \citep{Zhu2018}.
The differences between the top of the filled regions and the solid lines show the unique effect that superheavies have on the heating rates, differentiable from the actinides. 

For FRDM, there is no apparent difference between the ``$Z$ = all" and ``$Z<104$" cases owing to FRDM producing very small mass fraction of superheavy elements (see Figure \ref{fig:massfracs}).
With the HFB model, several discernible features appear on time scales ranging from about 0.2 hours out to several days.
Two of the most prominent superheavy-unique features appear at approximately seven hours ($10^{-0.5}$ days) and several days ($10^{0.4}$ days) post-merger.

The first feature in the heating rate at about seven hours is primarily due to the spontaneous fission of odd-$A$ nuclei with $Z=104$--106 and $N=169$--171.
The cause of this spontaneous fission dominance is two-fold; first, the abundance of material that is able to reach this region is due to the high fission barriers (i.e., lower fission probabilities) associated with the HFB model \citep{Vassh2019}. 
This effect is apparent in Figure \ref{fig:massfracs}, which shows, for example, the mass fraction of $Z=104$ isotopes that is several orders of magnitude larger in HFB than in the results from the FRDM calculation.
Secondly, the probabilities for nuclei to decay by fission, $\beta$, or $\alpha$ differ between FRDM and HFB, therefore affecting both the energy and time scales predicted between the two models.

The second peak in the HFB ``$Z$ = all" heating rate at a few days post-merger largely stems from the spontaneous fission of nuclei with $A=273$--277. Again, we attribute this to both the build-up of material at $Z=104$ as well as the theoretical rates (and, therefore, branching ratios) differing between FRDM and HFB that contribute to the population of these nuclei.
In particular, it is worth noting that other contemporary studies \citep[e.g.,][]{Giuliani2018,Kullmann2022} do not see the same build-up of material at $Z=104$ or its associated heating-rate excess.
This difference from literature comes from the employed $\beta$-decay rates.
In particular, \citet{Giuliani2018} and \citet{Kullmann2022} use rates from \citet{Marketin2016} in their \rp\ network calculations, while the present work uses $\beta$-decay rates recalculated using the nuclear masses of the nuclear model employed (i.e., HFB, FRDM, etc.) from \citet{Mumpower2015}.
It is known that the inclusion of first-forbidden transitions in the \citet{Marketin2016} $\beta$-decay rates lead them to be overall faster above the $N=126$ shell closure compared to those calculated by \citet{Mumpower2015}.
This difference leads to nearly negligible heating rates from spontaneous fission for calculations that use \citet{Marketin2016} $\beta$-decay rates, as shown in \citet{lund2023}.
For example, in the present calculation the (theoretical) half life of $^{275}$Rf is calculated to be much longer with HFB (7.1 hours) than with FRDM (20 minutes), controlling the overall time that its $\beta$-decay product---$^{275}$Db---takes to undergo spontaneous fission.
Therefore, it is not only fission rates that ultimately determine heating from the superheavies, but the heating is also moderated by the $\beta$ decay rates and the nuclear mass model.
The heating rates of individual superheavy elements at specific times are given in the Appendix.
In the next section, we investigate how this signature in the heating rate manifests as an observational signature in the light curve.

	\begin{figure*}[tbh]
	\centering
    \includegraphics[width=\columnwidth]{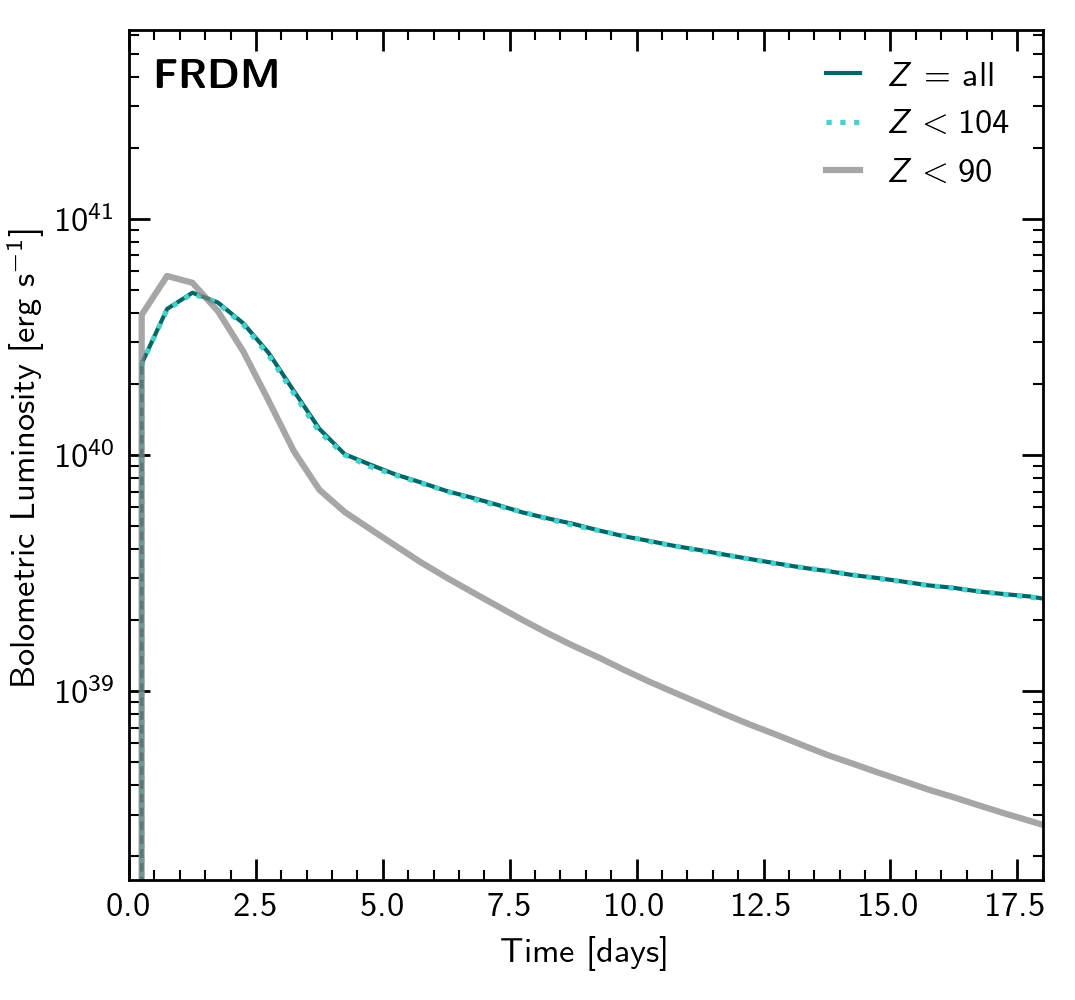}\qquad
    \includegraphics[width=\columnwidth]{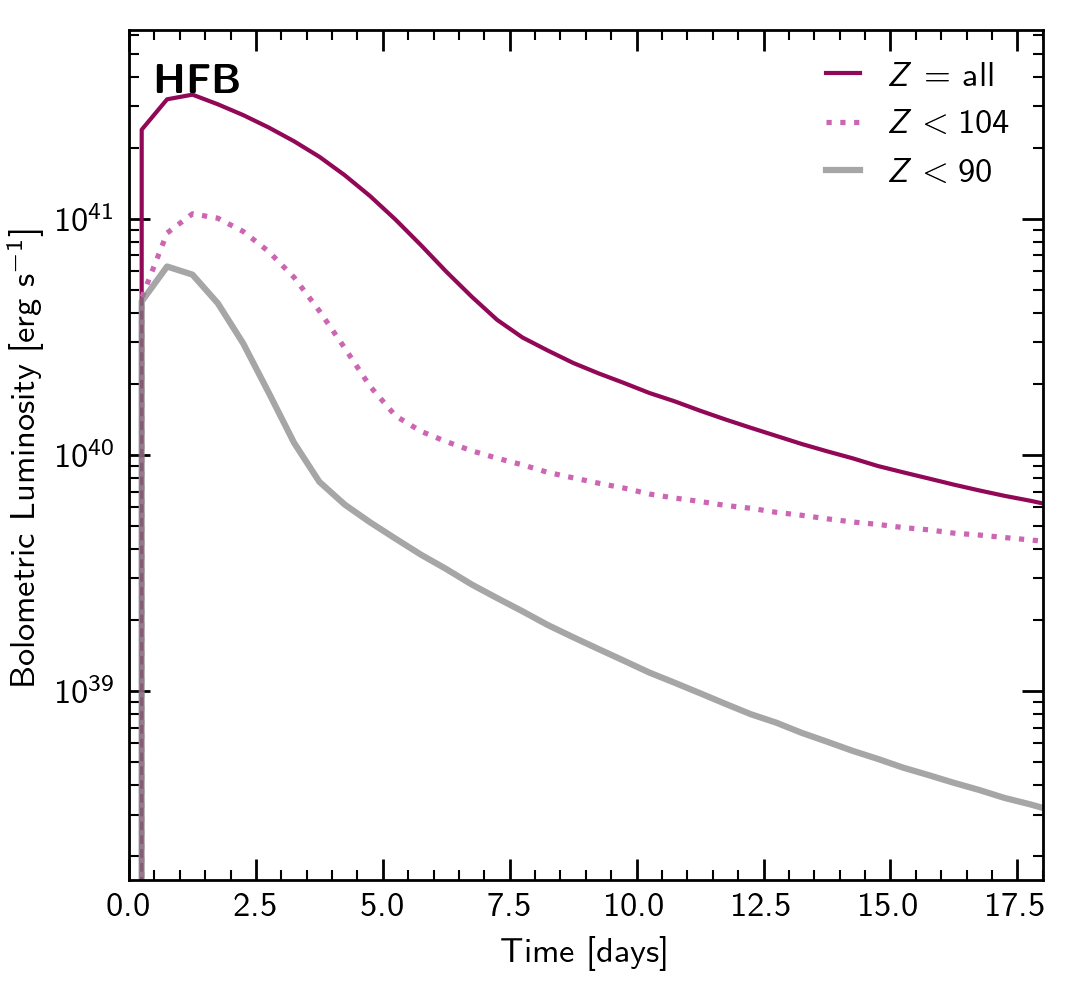}
	\caption{Bolometric light curves using the FRDM nuclear model (left) and the HFB model (right) assuming an ejecta mass of 0.005 M$_\odot$. For each nuclear model, three sets of heating rates are used: only $Z<90$ elements (solid gray), only $Z<104$ elements (dotted colors), and all elements (solid colors).
	The paucity of superheavy elements produced in the FRDM simulation explains why the green solid (``$Z$=all") and light-blue dotted (``$Z<104$") curves appear identical; there is effectively no difference between these models.
    }
	\label{fig:lbol}
	\end{figure*}

\subsection{Thermalization and radiation transport}

To calculate light curves, we use both the PRISM composition and the predicted
fraction of the radioactive energy released by $\alpha$-decay, $\beta$-decay, and fission 
(the relevant decay processes on kilonova timescales).
With the energy partitioned by decay channel, we employ analytic estimates of particle-specific thermalization efficiencies \citep{Kasen2019} following \citet{zhu2021} to derive the rate at which thermal energy is imparted to the kilonova ejecta as a function of time.
This approach means that the difference in the thermalization rates of our models are due solely to differences in the relative importance of $\alpha$-day, $\beta$-decay, and fission;
we do not carry out a full calculation of particle emission and propagation \citep[though see][for a detailed discussion of self-consistent thermalization simulations]{barnes2021}.

In addition to different heating rates, the three variations we study here also have different compositions, which affects the light curves by altering the ejecta opacity.
When specifying the composition for each variant, we remove the indicated elements (i.e., actinides and superheavies, actinides only, or no elements) and renormalize the resulting composition so the mass fractions of the remaining elements sum to unity. 

The calculation of the opacity depends on the atomic structure of the elements in the ejecta.
Specifically, due to the periodicity of the periodic table, the atomic structure of elements is a function primarily of the block in the table that it occupies, and secondarily on its position within its row \citep{kasen2013_opacs, tanaka2013_opac, tanaka2020_opacs}.
More recent work \citep{Fontes2020,tanaka2020_opacs,Fontes2023} confirmed these trends.
Less clear is how the opacity depends on the principle quantum number within a given block of the periodic table.

Because we lack synthetic atomic data for the full set of elements synthesized by the \emph{r}-process, we use a simplified composition in our radiation transport simulations.
As $f$-block elements, lanthanides and actinides supply most of the opacity, and the kilonova emission is most sensitive to their total mass fraction and abundance pattern.
We take the lanthanide and actinide abundances from PRISM (or from the modified PRISM outputs), and modify them only to account for our lack of atomic data for $Z=72$ and the actinide species.
The abundance predicted for $Z=72$ is assigned to $Z=71$ in our proxy composition.
Additionally, the abundance predicted for each actinide element is assigned to the lanthanide that occupies the same position in the periodic table row, e.g., lanthanum is a proxy for actinium.
We model the contribution of $d$-block elements by summing their mass fractions and dividing the total evenly among $21 \leq Z \leq 28$. 
The remaining mass is assigned to $Z=20$, which acts as a low-opacity filler.
On the periodic table, the superheavy elements with $104\leq Z\leq 112$ occupy the $d$-block, and will therefore contribute as lower-opacity elements.
This simplification is necessary because the atomic structure of superheavy elements has not yet been studied.
Extensive studies of elemental opacities over a broad range of $Z$ \citep{tanaka2020_opacs} suggests that the dependence of opacity on principle quantum number is not overly strong.
However, studies of superheavy elements' atomic structure are needed to clarify the contribution of these species to the total opacity.
The detailed radiation transport is then calculated with \texttt{Sedona} \citep{kasen2006_rt,roth2015_rt} as in \citet{barnes2021}, assuming a spherically symmetric outflow with variable ejecta mass (see Section \ref{subsec:shelcs-varmej}) and an average ejecta velocity $v/c=0.1$.

\subsection{Light curves}

Figure \ref{fig:lbol} shows the bolometric light curves for two nuclear cases (FRDM and HFB) with the three heating-rate variations.
For all cases in Figure \ref{fig:lbol}, we assume an ejecta mass of 0.005 M$_\odot$.
The gray curves show the light curve with only $Z<90$ nuclei, that is, without any actinides or superheavy elements directly contributing to the heating rate.
The dotted colored curves show the light curve when actinides are allowed to contribute to the total heating; i.e., heating is from all nuclei with $Z<104$.
The sharp transition at a few days is when the ejecta goes from optically thick to optically thin, after which the bolometric luminosity simply traces the energy input.
Because local thermodynamic equilibrium is a poor model of the gas state in this regime, the colors predictions after the transitions are less reliable than those corresponding to the light-curve peak.
Already differences arise compared to the no-actinide case.
First more energy is radiated over the course of the light curve.
For FRDM, this manifests on the light-curve tail, which is much brighter when actinides are included in the heating.
For HFB, both the peak and the tail  of the ``$Z < 104$'' light curve are more luminous than for the ``$Z < 90$'' case.
The dramatic increase in brightness seen with HFB can be understood as a direct consequence of the order-of-magnitude greater heating rates in the HFB ``$Z$=all" case compared to the other cases.
Regardless of nuclear mass model, the light curve declines more slowly when actinides are present.
Second, the added opacity of the actinides also broadens the light curve, delaying the peak relative to models without actinides.
The solid, colored curves in Figure \ref{fig:lbol} show the bolometric luminosity achieved  when all nuclei (including the superheavies) participate in the heating.
For FRDM, the effect is negligible, owing to the paucity of superheavy elements that are produced when using FRLDM fission barriers (see Figure \ref{fig:massfracs}).
However, the higher fission barrier heights of HFB14 allow significant superheavy-element production, dramatically increasing the brightness and overall duration of the light curve.

Under the same astrophysical conditions---not only in thermodynamic evolution, but also in ejecta mass and velocity---the superheavy elements can leave a distinct signature on the light curve that is unique from the actinides.
If an event were to display such an evolution of its light, not only would it be the first observational evidence that superheavy elements are produced astrophysically, but it would also help constrain the unknown nuclear physics by placing limits on fission barrier heights of neutron-rich nuclei.

\subsection{Observationally differentiating superheavy-element production}\label{subsec:shelcs-varmej}

While the superheavy element signature is evident in Figure \ref{fig:lbol}, when observing kilonovae we do not know the ejecta mass a priori. 
The question remains whether there is a discernible signature of superheavy-element production in light curves that have similar time-dependent bolometric luminosities.
To explore this idea, we adjust the ejecta masses for each case such that the total heating rates at one day are equal to the FRDM ``$Z$ = all" case with 0.005 M$_\odot$ (see Figure \ref{fig:lbol}).
This normalization produces light curves of similar bolometric luminosities.
The HFB ejecta masses required to produce the desired heating rates are 0.00051, 0.00163, and 0.00541 M$_\odot$ for the ``$Z$ = all," ``$Z<104$," and ``$Z<90$" cases, respectively.
With these normalized ejecta masses, we recalculate the bolometric luminosities.

\begin{figure}[t]
	\centering
    \includegraphics[width=\columnwidth]{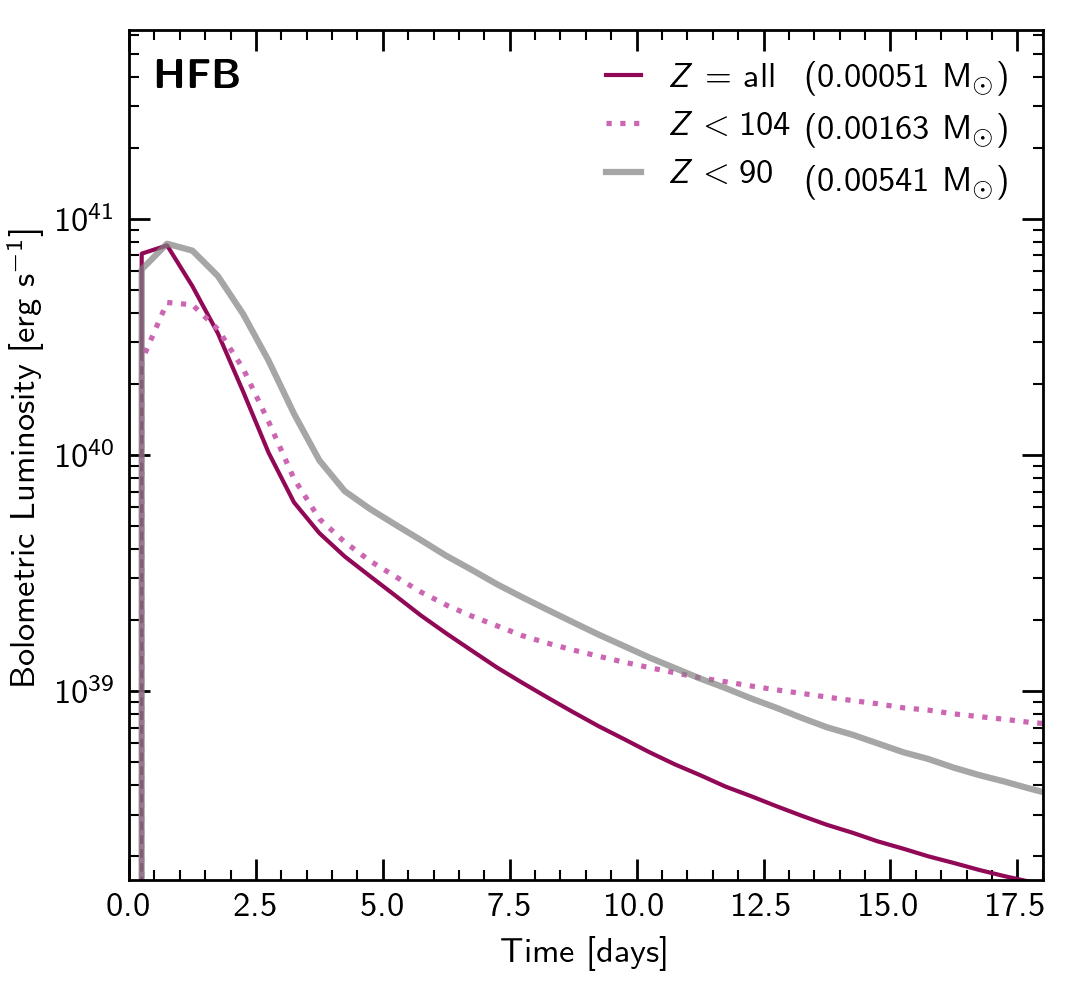}
	\caption{Bolometric light curves for the HFB model (right) with ejecta es renormalized such that the heating rates at one day is equal for all models. Three sets of heating rates: only $Z<90$ elements (gray), only $Z<104$ elements (dashed pink), and all elements (dark pink).}
	\label{fig:lbol_norm}
	\end{figure}

Figure \ref{fig:lbol_norm} shows the results for HFB when the ejecta masses are renormalized, corresponding to the case in which the light curves are roughly the same brightness at about one day.
Because the overall contribution of the actinides and superheavies is decreased (by virtue of lowering the total ejecta masses in those cases), their effects on the light curve are significantly less than in Figure \ref{fig:lbol}, in which identical ejecta masses were used for each light curve calculation.
Notably, the inclusion of actinides produces a light curve with a slightly fainter luminosity at peak and a slower-decaying evolution at late times (longer than five days).
However, if the composition includes the superheavies in addition to the actinides, the effect on the late-time light curve diminishes.
In addition, within the first few days, the light curve decays much more quickly (note how the dark pink peak is narrower than the other cases).

This effect arises both from the effective redistribution (between mass models) of the composition from high-opacity elements into low-opacity ones as well as the quick decay of those elements.
The superheavies act as additional $d$-block elements, optically similar to the iron-peak elements that could be expected to form in lower neutron-richness environments (such as an accretion disk wind), despite requiring a very high neutron flux in order to synthesize them.
This effect may be exacerbated by the excess heat from the decay of the superheavy elements, which can produce low-opacity ionization states in the ejecta at early times \citep{tanaka2020_opacs,barnes2021}.

	\begin{figure*}[ht]
	\centering
    \includegraphics[width=\columnwidth]{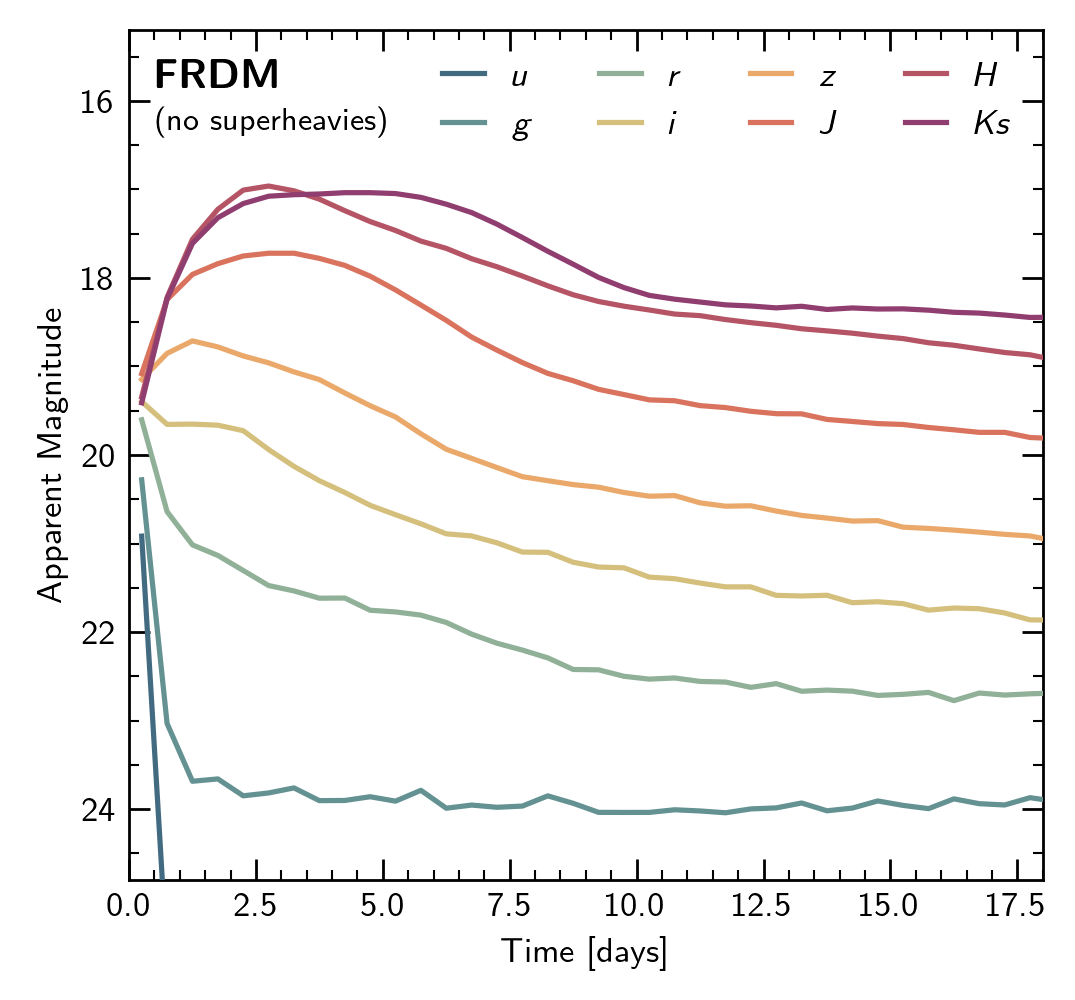}\qquad
	\includegraphics[width=\columnwidth]{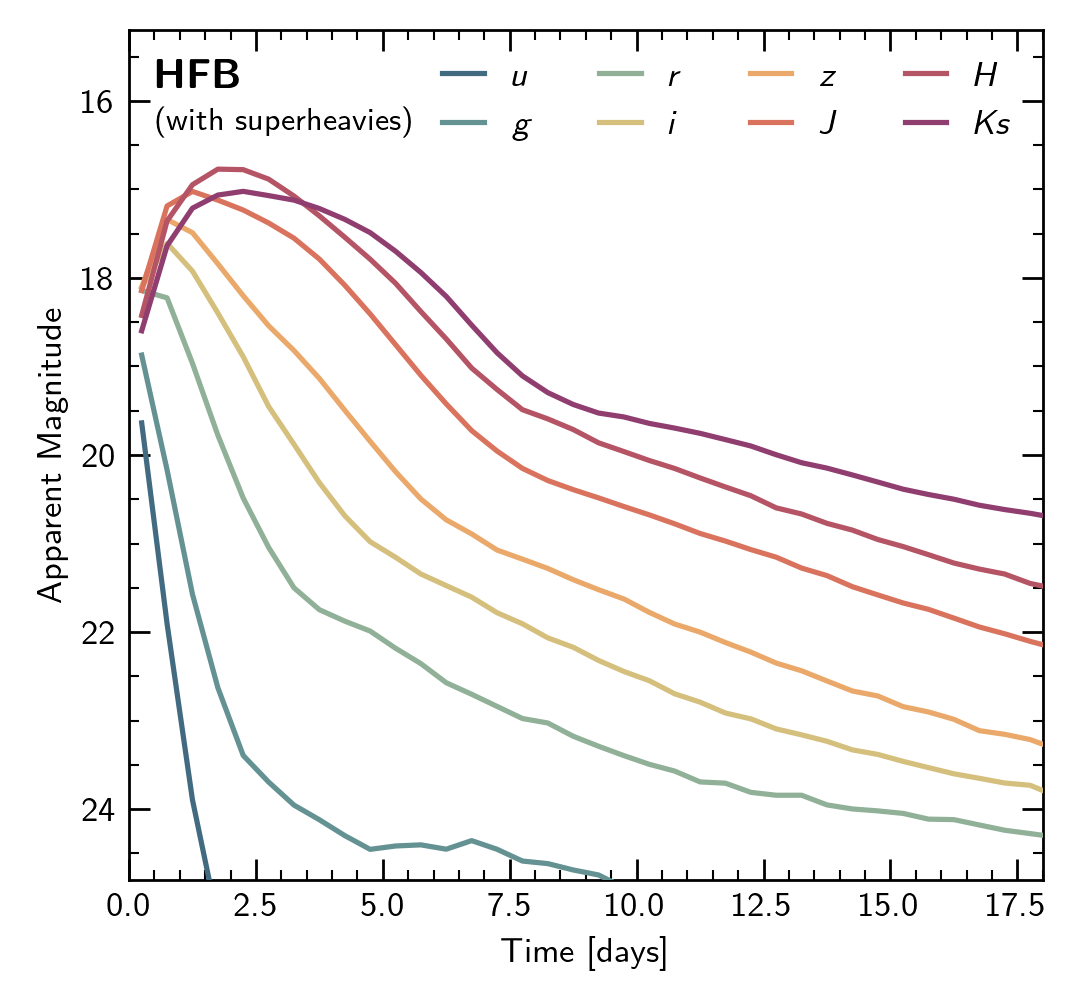}
    \caption{The effect of superheavy elements on broad-band light curves for the full FRDM (no superheavy elements) and HFB (with superheavy elements) under similar observational conditions (lines).
	}
	\label{fig:lbb}
	\end{figure*}

Figure \ref{fig:lbb} shows the broad band light curves in the normalized-mass case for both the FRDM and the HFB calculations in which the full range of elements are allowed to contribute to the heating rate.
Here we treat the FRDM calculation as the case in which superheavy elements are not present in the ejecta rather than manually removing the superheavy elements from the HFB calculation. 
Differences in these two light curves highlight the unique effects---both direct and indirect---that superheavy (and not just actinide) elements have on the kilonova light curve evolution.
The presence of actinides and lanthanides (and their associated heating) in the FRDM case leads to late-time (i.e., after the luminosity peak at a few days) emission in the red bands, while the fast and energetic decays of the superheavy elements in the HFB case lead to earlier, brighter emission in the blue bands.
However, neither start as blue-band bright as AT2017gfo.
Note that actinides are still present in the HFB calculation, and we stress that a component containing superheavy elements may only be one piece of real kilonova ejecta.
For example, the AT2017gfo data is commonly fit with two components \cite[see, e.g.,][]{cowperthwaite2017}: a ``blue" (low opacity) and a ``red" (high opacity).
A superheavy-element component in our investigation acts like an additional blue component in composition, despite containing high-opacity elements.
Further studies should explore whether the AT2017gfo data shows evidence for a separate, superheavy-element-containing component or whether the superheavy-element signature would have been shrouded by the emission of less neutron-rich ejecta.

\section{Conclusion}

In this study, we have explored and selected a combination of nuclear data and astrophysical conditions that allows production of superheavy elements in order to examine their potential impact on observables (i.e., kilonova light curves).
So far, no evidence of the natural production of $Z\geq 104$ elements has been definitively found \cite[though see claims of superheavy decay products found in meterorites; references in][]{Holmbeck2023}.
If superheavy elements have a unique effect on light curves, it may be possible to infer their production in NSMs.
Such an identification would be the first observational evidence of superheavy elements being synthesized by nature.
This possibility is the motivation behind the present analysis.

First, we achieved superheavy-element production by using the nuclear masses and fission barrier heights of the HFB model and the neutron-rich tidal tails of NSM ejecta.
We find that superheavy elements can lead to light curves that decline quickly after peak brightness and continue to fall through the following weeks post-merger.
This rapid decline is in conflict with the extended late-time emission that the presence of lanthanides and actinides has on kilonova ejecta.
In addition, ejecta that includes actinides and superheavy elements produces a light curve similar in evolution to a ten-times-as-massive actinide-free event.
Our results demonstrate another natural degeneracy between NSM ejecta composition and mass, this time due to the presence (or lack) of superheavy elements in the merger ejecta.

We have only considered spherically symmetric outflows by the neutron-rich tidal ejecta in this work since those conditions can favorably produce the superheavy elements.
In a multi-component kilonova model, a lanthanide-poor component could plausibly produce similar broad-band light curves, further increasing the degeneracy between ejecta mass and composition.
Whether the superheavy element signal that we have reported on here is observable will depend on future 3D modeling efforts. There still remains significant uncertainty in the three-dimensional distribution of compact object merger ejecta and what fraction of the total ejecta mass is capable of reaching the high neutron-richness required for superheavy-element formation.
Furthermore, the ejecta can sensitively depend on the properties of the compact objects that are merging, which will vary case-by-case.
Although there is general conjecture that neutron star mergers eject the majority of their mass through lanthanide-poor disk winds, simulations that run out to \mbox{$\mathcal O$(seconds)} find that viscous properties of the accretion disk dominate over neutrino interactions, causing a delayed ejecta that is instead very neutron-rich \cite[e.g.,][]{Fernandez2013}.
In addition, even a neutron-rich component that is subdominant may contribute significantly to the light curve under a favorable viewing angle \cite[e.g.,][]{Fujibayashi2018}.
Full, three-dimensional neutron-star merger models that include high-fidelity neutrino-transport calculations---especially out to late times ($\mathcal O$(seconds))---will help reveal the total possible mass fraction of superheavy elements in compact object merger ejecta.

The lack of an extended red light curve in the HFB case in Figure~\ref{fig:lbb} does not rule out the presence of actinides in merger ejecta, and care should be taken when deriving lanthanide mass-fractions from kilonova observations.
If there is a degeneracy in the light curve evolution---a photometric quantity---then one possibility that remains to be explored is whether the superheavy elements are distinguishable spectroscopically.
Such a study would require extensive atomic physics dedicated to characterizing the features that could be visible in kilonova spectra.
Both through spectroscopic and photometric signatures, this work reinforces the need for contemporary atomic modeling in the superheavy-element regime.

The way in which superheavies contribute to the total heating rate of NSM ejecta depends on fission rates at high atomic mass numbers.
Therefore, the possibility of the astrophysical production of superheavy elements is still contingent on as-of-yet theoretical nuclear data.
For example, the feature in the heating rate identified between several hours and days post-merger relies on the $\beta$ decay rates of heavy actinides and superheavy elements ($A=273$--277) being comparably long. These rates are currently entirely theoretical. The real rates being much shorter or much longer will have corresponding observational consequences on the light curve.
Advances in nuclear theory and experiment are critical to addressing nuclear unknowns, including the decay rates of heavy nuclei and whether fission suppresses superheavy-element production in NSMs.
Currently, only the N=126 Factory at Argonne National Laboratory is able to probe the heaviest elements terrestrially.
Kilonovae offer an astrophysical way of investigating nuclear properties; the presence (or lack) of a superheavy-element signature can itself offer constraints on the nuclear physics of very heavy nuclei.

With the possibility of more observational light curves accompanying NSM candidates detected by the next gravitational wave observing run, we have the new opportunity to observe---for the first time---superheavy-element formation as it occurs.
However, there is a growing need to resolve discrepancies and degeneracies from nuclear and astrophysics modeling before definitive conclusions about elemental production are drawn from kilonova observations.


\begin{acknowledgements}

E.M.H.\ acknowledges support for this work provided by NASA through the NASA Hubble Fellowship grant HST-HF2-51481.001 awarded by the Space Telescope Science Institute, which is operated by the Association of Universities for Research in Astronomy, Inc., for NASA, under contract NAS5-26555.
J.B.\ gratefully acknowledges support from the Gordon and Betty Moore Foundation through Grant GBMF5076 and the National Science Foundation under Grant No. NSF PHY-1748958.
K.A.L., T.M.S., and M.R.M.\ were supported by the U.S.\ Department of Energy (DOE) through the Los Alamos National Laboratory (LANL). LANL is operated by Triad National Security, LLC, for the National Nuclear Security Administration of U.S.\ Department of Energy (Contract No.\ 89233218CNA000001). 
G.C.M.\ acknowledges support from the NSF (N3AS PFC) grant No. PHY-2020275, as well as from U.S.\ DOE contract No.\ DEFG0202ER41216.
K.A.L.\ and G.C.M.\ acknowledge support from the Fission in r-Process Elements (FIRE) topical collaboration in nuclear theory, funded by the U.S.\ DOE contract No.\ DEAC5207NA27344.
K.A.L.\ gratefully acknowledges the support of the U.S.\ DOE through the LANL/LDRD Program and the Center for Non Linear Studies for this work.
This paper is approved for unlimited release, assigned LA-UR-23-23123.
\end{acknowledgements}


\bibliographystyle{aasjournal}
\bibliography{bibliography}

\appendix
\section{Data}

Here we supply the effective heating rates at select times in our calculation.
The effective heating rates are calculated as in Equation 3 of \citet{lund2023}:
    \begin{equation}
        \dot Q(t) = \sum_i \dot q_i(t)\, f_i(M_{\text{ej}}, v_{\text{ej}}, t)\, M_{\text{ej}},
    \end{equation}
where $M_{\text{ej}}$ is 0.005~M$_\odot$, and $v_{\text{ej}}=0.1\,c$.
For each nucleus, the effective heating rate is calculated for each reaction/decay channel: $\beta$-decay, $\alpha$-decay, and fission (the sum of the three individual mechanisms: spontaneous fission, $\beta$-delayed fission, and neutron-induced fission).
At any particular time, the decay through a particular channel will constitute the majority of the heat from that nucleus.

Tables \ref{tab:hr_hours}--\ref{tab:hr_month} show the mass fractions of superheavy nuclei at particular times: 7.5 hours, 1 day, 1 week, and 1 month.
The effective heating rate from the channel contributing to the majority of the heating rate is shown for each nucleus, as well as to which channel each value is attributed.
The values for the HFB and FRDM calculations are both shown; a blank value indicates that the nucleus does not exist (or exists in an abundance below a computational threshold value) at that particular timestep.

\input{data_7hours}
\input{data_1day}
\input{data_1week}
\input{data_1month}

\end{document}

%% file: data_7hours.tex
\startlongtable
\begin{deluxetable*}{c c c c c}
\tablecaption{Mass fractions ($X$) and effective heating rates of the decay channel contributing the majority of instantaneous heat ($\dot Q_{\text{max}}$) for each of the superheavy nuclei at 7.5 hours. The channel responsible is indicated in parentheses: ``f" for the fission channels, ``$\alpha$" for $\alpha$-decay, and ``$\beta$" for $\beta$-decay. Values are shown for both HFB and FRDM.\label{tab:hr_hours}}
\tablehead{
\colhead{Nucleus} & \colhead{$X^{\text{HFB}}$} & \colhead{$\dot{Q}^{\text{HFB}}_{\text{max}}$ (channel)} &
\colhead{$X^{\text{FRDM}}$} & \colhead{$\dot{Q}^{\text{FRDM}}_{\text{max}}$ (channel)}}
\startdata
$^{267}$Rf &  &  & 6.56$\times$10$^{-7}$ & 8.93$\times$10$^{38}$ (f) \\
$^{268}$Rf & 9.36$\times$10$^{-7}$ & 6.69$\times$10$^{36}$ ($\alpha$) & 2.79$\times$10$^{-10}$ & 1.18$\times$10$^{32}$ ($\alpha$) \\
$^{269}$Rf & 9.89$\times$10$^{-3}$ & 1.41$\times$10$^{39}$ (f) & 6.08$\times$10$^{-6}$ & 6.18$\times$10$^{36}$ ($\alpha$) \\
$^{270}$Rf & 2.01$\times$10$^{-3}$ & 2.31$\times$10$^{41}$ (f) & 3.34$\times$10$^{-6}$ & 3.83$\times$10$^{38}$ (f) \\
$^{271}$Rf & 9.51$\times$10$^{-3}$ & 3.17$\times$10$^{41}$ (f) & 6.61$\times$10$^{-10}$ & 2.20$\times$10$^{34}$ (f) \\
$^{272}$Rf & 6.44$\times$10$^{-9}$ & 1.07$\times$10$^{38}$ (f) & 6.95$\times$10$^{-12}$ & 1.16$\times$10$^{35}$ (f) \\
$^{273}$Rf & 1.98$\times$10$^{-3}$ & 2.60$\times$10$^{42}$ (f) & 1.19$\times$10$^{-6}$ & 1.56$\times$10$^{39}$ (f) \\
$^{274}$Rf & 2.27$\times$10$^{-11}$ & 3.30$\times$10$^{36}$ (f) &  &  \\
$^{275}$Rf & 5.36$\times$10$^{-3}$ & 1.41$\times$10$^{42}$ (f) & 5.45$\times$10$^{-12}$ & 1.43$\times$10$^{33}$ (f) \\
$^{276}$Rf & 1.13$\times$10$^{-10}$ & 3.12$\times$10$^{35}$ (f) & 1.47$\times$10$^{-8}$ & 4.04$\times$10$^{37}$ (f) \\
$^{277}$Rf & 1.06$\times$10$^{-5}$ & 9.20$\times$10$^{37}$ ($\beta$) &  & \\
$^{278}$Rf & 9.24$\times$10$^{-12}$ & 5.06$\times$10$^{32}$ (f) & 1.37$\times$10$^{-9}$ & 7.49$\times$10$^{34}$ (f) \\
$^{279}$Rf & 3.61$\times$10$^{-9}$ & 9.01$\times$10$^{34}$ ($\beta$) &  &  \\
$^{269}$Db & 5.91$\times$10$^{-7}$ & 4.74$\times$10$^{38}$ ($\alpha$) &  &  \\
$^{271}$Db &  &  & 3.64$\times$10$^{-6}$ & 5.89$\times$10$^{37}$ ($\alpha$) \\
$^{272}$Db &  &  & 1.14$\times$10$^{-12}$ & 1.26$\times$10$^{32}$ (f) \\
$^{273}$Db & 2.65$\times$10$^{-4}$ & 1.22$\times$10$^{41}$ (f) & 9.51$\times$10$^{-7}$ & 4.37$\times$10$^{38}$ (f) \\
$^{275}$Db & 9.90$\times$10$^{-5}$ & 2.34$\times$10$^{42}$ (f) & 4.64$\times$10$^{-12}$ & 1.09$\times$10$^{35}$ (f) \\
$^{276}$Db & 1.57$\times$10$^{-9}$ & 2.85$\times$10$^{37}$ (f) &  &  \\
$^{277}$Db & 2.88$\times$10$^{-5}$ & 3.14$\times$10$^{40}$ (f) & 3.30$\times$10$^{-9}$ & 3.48$\times$10$^{36}$ (f) \\
$^{278}$Db & 2.65$\times$10$^{-6}$ & 8.47$\times$10$^{39}$ (f) & 8.32$\times$10$^{-11}$ & 3.56$\times$10$^{34}$ (f) \\
$^{279}$Db & 9.55$\times$10$^{-9}$ & 3.14$\times$10$^{37}$ (f) &  &  \\
$^{280}$Db & 8.54$\times$10$^{-10}$ & 4.48$\times$10$^{36}$ (f) &  &  \\
$^{281}$Db & 3.45$\times$10$^{-10}$ & 1.27$\times$10$^{36}$ (f) &  &  \\
$^{273}$Sg & 4.27$\times$10$^{-7}$ & 1.83$\times$10$^{40}$ (f) & 6.32$\times$10$^{-9}$ & 2.71$\times$10$^{38}$ (f) \\
$^{275}$Sg & 1.83$\times$10$^{-4}$ & 6.58$\times$10$^{41}$ (f) & 2.92$\times$10$^{-9}$ & 1.05$\times$10$^{37}$ (f) \\
$^{276}$Sg & 1.42$\times$10$^{-12}$ & 1.05$\times$10$^{36}$ (f) &  &  \\
$^{277}$Sg & 7.82$\times$10$^{-6}$ & 6.03$\times$10$^{40}$ (f) & 1.83$\times$10$^{-9}$ & 1.41$\times$10$^{37}$ (f) \\
$^{278}$Sg & 7.15$\times$10$^{-9}$ & 1.36$\times$10$^{39}$ (f) & 9.57$\times$10$^{-12}$ & 1.82$\times$10$^{36}$ (f) \\
$^{279}$Sg & 1.16$\times$10$^{-4}$ & 4.54$\times$10$^{40}$ (f) & 1.86$\times$10$^{-9}$ & 7.28$\times$10$^{35}$ (f) \\
$^{280}$Sg & 2.27$\times$10$^{-9}$ & 1.53$\times$10$^{37}$ (f) &  &  \\
$^{281}$Sg & 1.70$\times$10$^{-6}$ & 5.00$\times$10$^{36}$ (f) &  &  \\
$^{282}$Sg & 9.35$\times$10$^{-10}$ & 3.10$\times$10$^{35}$ (f) &  &  \\
$^{283}$Sg & 6.28$\times$10$^{-12}$ & 2.87$\times$10$^{33}$ (f) &  &  \\
$^{277}$Bh & 5.39$\times$10$^{-10}$ & 5.20$\times$10$^{37}$ ($\alpha$) & 3.80$\times$10$^{-11}$ & 5.09$\times$10$^{35}$ (f) \\
$^{282}$Bh & 1.01$\times$10$^{-10}$ & 1.94$\times$10$^{35}$ (f) &  &  \\
$^{283}$Bh & 2.40$\times$10$^{-11}$ & 3.81$\times$10$^{33}$ (f) &  &  \\
$^{282}$Hs & 3.48$\times$10$^{-12}$ & 7.56$\times$10$^{33}$ (f) &  &  \\
$^{283}$Hs & 1.61$\times$10$^{-12}$ & 2.54$\times$10$^{31}$ ($\alpha$) &  & \\
\enddata
\end{deluxetable*}

%% file: data_1day.tex
\startlongtable
\begin{deluxetable*}{c c c c c}
\tablecaption{As in Table~\ref{tab:hr_hours}, but at one day.\label{tab:hr_day}}
\tablehead{
\colhead{Nucleus} & \colhead{$X^{\text{HFB}}$} & \colhead{$\dot{Q}^{\text{HFB}}_{\text{max}}$ (channel)} &
\colhead{$X^{\text{FRDM}}$} & \colhead{$\dot{Q}^{\text{FRDM}}_{\text{max}}$ (channel)}}
\startdata
$^{267}$Rf &  &  & 7.82$\times$10$^{-7}$ & 1.06$\times$10$^{39}$ (f) \\
$^{268}$Rf & 3.02$\times$10$^{-7}$ & 2.13$\times$10$^{36}$ ($\alpha$) & 2.66$\times$10$^{-10}$ & 1.12$\times$10$^{32}$ ($\alpha$) \\
$^{269}$Rf & 1.02$\times$10$^{-2}$ & 1.44$\times$10$^{39}$ (f) & 5.44$\times$10$^{-6}$ & 5.47$\times$10$^{36}$ ($\alpha$) \\
$^{270}$Rf & 1.17$\times$10$^{-3}$ & 1.33$\times$10$^{41}$ (f) & 1.94$\times$10$^{-6}$ & 2.21$\times$10$^{38}$ (f) \\
$^{271}$Rf & 7.63$\times$10$^{-3}$ & 2.53$\times$10$^{41}$ (f) &  &  \\
$^{273}$Rf & 5.57$\times$10$^{-6}$ & 7.30$\times$10$^{39}$ (f) & 1.02$\times$10$^{-8}$ & 1.34$\times$10$^{37}$ (f) \\
$^{275}$Rf & 1.73$\times$10$^{-4}$ & 4.51$\times$10$^{40}$ (f) &  &  \\
$^{276}$Rf &  &  & 4.96$\times$10$^{-12}$ & 1.36$\times$10$^{34}$ (f) \\
$^{277}$Rf & 1.04$\times$10$^{-10}$ & 6.37$\times$10$^{32}$ (f) &  &  \\
$^{278}$Rf &  &  & 2.67$\times$10$^{-11}$ & 1.46$\times$10$^{33}$ (f) \\
$^{269}$Db & 6.11$\times$10$^{-7}$ & 4.83$\times$10$^{38}$ ($\alpha$) &  &  \\
$^{271}$Db &  &  & 6.72$\times$10$^{-7}$ & 1.08$\times$10$^{37}$ ($\alpha$) \\
$^{273}$Db & 3.50$\times$10$^{-5}$ & 1.60$\times$10$^{40}$ (f) & 1.42$\times$10$^{-7}$ & 6.50$\times$10$^{37}$ (f) \\
$^{275}$Db & 3.19$\times$10$^{-6}$ & 7.50$\times$10$^{40}$ (f) &  &  \\
$^{277}$Db & 9.38$\times$10$^{-11}$ & 1.02$\times$10$^{35}$ (f) &  & \\
$^{278}$Db & 1.91$\times$10$^{-12}$ & 6.08$\times$10$^{33}$ (f) & 1.62$\times$10$^{-12}$ & 6.93$\times$10$^{32}$ (f) \\
$^{273}$Sg & 5.70$\times$10$^{-8}$ & 2.43$\times$10$^{39}$ (f) & 9.50$\times$10$^{-10}$ & 4.06$\times$10$^{37}$ (f) \\
$^{275}$Sg & 6.06$\times$10$^{-6}$ & 2.17$\times$10$^{40}$ (f) &  & \\
$^{277}$Sg & 2.14$\times$10$^{-11}$ & 1.64$\times$10$^{35}$ (f) &  & \\
$^{279}$Sg & 1.52$\times$10$^{-5}$ & 5.91$\times$10$^{39}$ (f) & 5.54$\times$10$^{-10}$ & 2.16$\times$10$^{35}$ (f) \\
$^{281}$Sg & 1.68$\times$10$^{-6}$ & 4.90$\times$10$^{36}$ (f) &  &  \\
$^{282}$Sg & 8.36$\times$10$^{-11}$ & 2.76$\times$10$^{34}$ (f) &  &  \\
$^{282}$Bh & 1.02$\times$10$^{-11}$ & 1.95$\times$10$^{34}$ (f) &  &  \\
$^{283}$Bh & 9.81$\times$10$^{-12}$ & 1.55$\times$10$^{33}$ (f) &  &  \\
$^{282}$Hs & 3.69$\times$10$^{-13}$ & 7.97$\times$10$^{32}$ (f) &  &  \\
$^{283}$Hs & 2.15$\times$10$^{-12}$ & 3.35$\times$10$^{31}$ ($\alpha$) &  &  \\
\enddata
\end{deluxetable*}

%% file: data_1week.tex
\startlongtable
\begin{deluxetable*}{c c c c c}
\tablecaption{As in Table~\ref{tab:hr_hours}, but at one week.\label{tab:hr_week}}
\tablehead{
\colhead{Nucleus} & \colhead{$X^{\text{HFB}}$} & \colhead{$\dot{Q}^{\text{HFB}}_{\text{max}}$ (channel)} &
\colhead{$X^{\text{FRDM}}$} & \colhead{$\dot{Q}^{\text{FRDM}}_{\text{max}}$ (channel)}}
\startdata
$^{267}$Rf &  &  & 9.97$\times$10$^{-8}$ & 1.31$\times$10$^{38}$ (f) \\
$^{268}$Rf & 6.28$\times$10$^{-10}$ & 4.08$\times$10$^{33}$ ($\alpha$) & 1.73$\times$10$^{-10}$ & 6.71$\times$10$^{31}$ ($\alpha$) \\
$^{269}$Rf & 9.65$\times$10$^{-3}$ & 1.32$\times$10$^{39}$ (f) & 2.05$\times$10$^{-6}$ & 1.90$\times$10$^{36}$ ($\alpha$) \\
$^{270}$Rf & 5.70$\times$10$^{-5}$ & 6.33$\times$10$^{39}$ (f) & 1.74$\times$10$^{-8}$ & 1.93$\times$10$^{36}$ (f) \\
$^{271}$Rf & 2.20$\times$10$^{-3}$ & 7.10$\times$10$^{40}$ (f) &  &  \\
$^{273}$Rf & 1.02$\times$10$^{-10}$ & 1.30$\times$10$^{35}$ (f) &  &  \\
$^{275}$Rf & 7.35$\times$10$^{-11}$ & 1.87$\times$10$^{34}$ (f) &  &  \\
$^{277}$Rf & 9.13$\times$10$^{-11}$ & 5.46$\times$10$^{32}$ (f) &  &  \\
$^{269}$Db & 5.79$\times$10$^{-7}$ & 4.23$\times$10$^{38}$ ($\alpha$) &  &  \\
$^{271}$Db &  &  & 1.05$\times$10$^{-12}$ & 1.54$\times$10$^{31}$ ($\alpha$) \\
$^{273}$Db & 1.38$\times$10$^{-10}$ & 6.15$\times$10$^{34}$ (f) &  &  \\
$^{275}$Db & 1.34$\times$10$^{-12}$ & 3.07$\times$10$^{34}$ (f) &  &  \\
$^{277}$Db & 7.18$\times$10$^{-11}$ & 7.57$\times$10$^{34}$ (f) &  &  \\
$^{275}$Sg & 2.31$\times$10$^{-12}$ & 8.03$\times$10$^{33}$ (f) &  &  \\
$^{277}$Sg & 1.56$\times$10$^{-11}$ & 1.16$\times$10$^{35}$ (f) &  &  \\
$^{279}$Sg & 3.79$\times$10$^{-10}$ & 1.44$\times$10$^{35}$ (f) &  &  \\
$^{281}$Sg & 1.54$\times$10$^{-6}$ & 4.38$\times$10$^{36}$ (f) &  & 
\enddata
\end{deluxetable*}

%% file: data_1month.tex
\startlongtable
\begin{deluxetable*}{c c r c c r c}
\tablecaption{As in Table~\ref{tab:hr_hours}, but at one month.\label{tab:hr_month}}
\tablehead{
\colhead{Nucleus} & \colhead{$X^{\text{HFB}}$} & \colhead{$\dot{Q}^{\text{HFB}}_{\text{max}}$ (channel)} &
\colhead{$X^{\text{FRDM}}$} & \colhead{$\dot{Q}^{\text{FRDM}}_{\text{max}}$ (channel)}}
\startdata
$^{267}$Rf &  &  & 3.68$\times$10$^{-11}$ & 4.38$\times$10$^{34}$ (f) \\
$^{268}$Rf &  &  & 3.30$\times$10$^{-11}$ & 9.56$\times$10$^{30}$ ($\alpha$) \\
$^{269}$Rf & 7.79$\times$10$^{-3}$ & 9.65$\times$10$^{38}$ (f) & 4.95$\times$10$^{-8}$ & 3.44$\times$10$^{34}$ ($\alpha$) \\
$^{270}$Rf & 9.14$\times$10$^{-10}$ & 9.17$\times$10$^{34}$ (f) &  & \\
$^{271}$Rf & 1.87$\times$10$^{-5}$ & 5.46$\times$10$^{38}$ (f) &  & \\
$^{273}$Rf & 7.28$\times$10$^{-11}$ & 8.39$\times$10$^{34}$ (f) &  & \\
$^{277}$Rf & 6.53$\times$10$^{-11}$ & 3.53$\times$10$^{32}$ (f) &  & \\
$^{269}$Db & 4.67$\times$10$^{-7}$ & 2.57$\times$10$^{38}$ ($\alpha$) &  & \\
$^{273}$Db & 1.19$\times$10$^{-11}$ & 4.81$\times$10$^{33}$ (f) &  & \\
$^{277}$Db & 5.13$\times$10$^{-11}$ & 4.89$\times$10$^{34}$ (f) &  & \\
$^{277}$Sg & 1.12$\times$10$^{-11}$ & 7.53$\times$10$^{34}$ (f) &  & \\
$^{281}$Sg & 1.10$\times$10$^{-6}$ & 2.83$\times$10$^{36}$ (f) &  & \\
\enddata
\end{deluxetable*}